\documentclass[a4paper]{amsart}

\usepackage{ifxetex}
\ifxetex
\else
\usepackage[utf8]{inputenc}
\fi

\usepackage{amsmath}
\usepackage{mathcomp,amsfonts,amssymb}
\usepackage[mathscr]{euscript}
\usepackage{algorithmic,algorithm}
\usepackage{paralist}
\usepackage{graphicx}
\usepackage{placeins}

\usepackage{enumitem}
\usepackage{dcolumn}

\usepackage[numbers]{dmnatbib}

\usepackage{listings}
\usepackage{tikz}
\usetikzlibrary{arrows,automata,positioning}

\tikzstyle{state}=[circle,fill=black!25,minimum size=13pt,inner sep=0pt]
\tikzstyle{rstate}=[rectangle,fill=black!25,minimum size=13pt,inner sep=0pt]
\tikzstyle{transition}=[rectangle,semithick,draw=black!75,
  			  minimum size=4mm]
\tikzstyle{transition2}=[transition,rectangle,thick,dashed,
  			  minimum size=4mm]
\tikzstyle{PRstate}=[circle,double,draw,fill=blue!15,minimum size=13pt,inner sep=0pt]
\tikzstyle{polyhedra}=[blue!25,opacity=0.5,pattern=north west lines,pattern
color=blue]
\tikzstyle{line}=[black,thick]

\newcommand{\mytitle}{Succinct Representations for Abstract Interpretation}
\usepackage[pdftitle={\mytitle},pdfauthor={Julien Henry, David Monniaux, Matthieu Moy}]{hyperref}

\newcommand{\abstr}[1]{#1^\sharp}
\newcommand{\parts}[1]{\mathscr{P}(#1)}
\newcommand{\ZZ}{\mathbb{Z}}
\newcommand{\QQ}{\mathbb{Q}}
\newcommand{\RR}{\mathbb{R}}
\newcommand{\widening}{\mathop{\triangledown}}

\graphicspath{{figures/}{gnuplot/}}

\usepackage{ifthen}
\newboolean{draft}
\setboolean{draft}{false}

\ifthenelse{\boolean{draft}}{
  \newcommand{\avirer}[1]{{\color{red} #1}}
  \newcommand{\MM}[1]{{\color{blue} TODO(MM): #1}}
  \newcommand{\JH}[1]{{\color{teal} TODO(JH): #1}}
  \newcommand{\DM}[1]{{\color{violet} TODO(DM): #1}}
}{
  \newcommand{\avirer}[1]{}
  \newcommand{\MM}[1]{}
  \newcommand{\JH}[1]{}
  \newcommand{\DM}[1]{}
}

\begin{document}
\title[Succinct representations for abstract interpretation]{Succinct representations for abstract interpretation \\Combined analysis algorithms and experimental evaluation}
\thanks{\raisebox{-0.5em}{\includegraphics[height=1.5em]{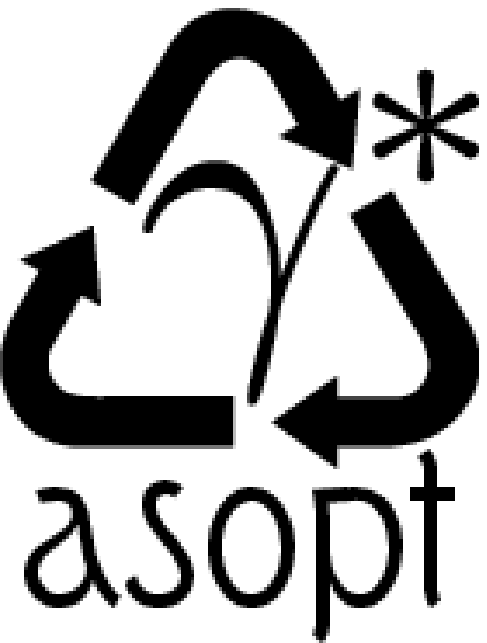}} This work was partially funded by ANR project ``ASOPT''}

\author{Julien Henry}
\address{J. Henry\\VERIMAG\\2 av de Vignate\\38610 Gières\\France}
\email{Julien.Henry@imag.fr}
\urladdr{http://www-verimag.imag.fr/~jhenry/}
\thanks{Julien Henry is a graduate student at Université Joseph Fourier, VERIMAG laboratory. VERIMAG is a joint laboratory of Université Joseph Fourier, CNRS and Grenoble-INP}

\author{David Monniaux}
\address{D. Monniaux\\VERIMAG\\2 av de Vignate\\38610 Gières\\France}
\email{David.Monniaux@imag.fr}
\urladdr{http://www-verimag.imag.fr/~monniaux/}
\thanks{David Monniaux is researcher at CNRS, VERIMAG laboratory}

\author{Matthieu Moy}
\address{M. Moy\\VERIMAG\\2 av de Vignate\\38610 Gières\\France}
\email{Matthieu.Moy@grenoble-inp.fr}
\urladdr{http://www-verimag.imag.fr/~moy/}
\thanks{Matthieu Moy is assistant professor at Grenoble-INP, VERIMAG laboratory}

\begin{abstract}
Abstract interpretation techniques can be made more precise by distinguishing
paths inside loops, at the expense of possibly exponential complexity. SMT-solving techniques and sparse representations of paths and sets of paths avoid this pitfall.

We improve previously proposed techniques for guided static analysis and the generation of disjunctive invariants by combining them with techniques for succinct representations of paths and symbolic representations for transitions based on static single assignment.

Because of the non-monotonicity of the results of abstract interpretation with widening operators, it is difficult to conclude that some abstraction is more precise than another based on theoretical local precision results. We thus conducted extensive comparisons between our new techniques and previous ones, on a variety of open-source packages.
\end{abstract}

\maketitle

\section{Introduction}
Static analysis by abstract interpretation is a fully automatic program analysis method. When applied to imperative programs, it computes an inductive invariant mapping each program location (or a subset thereof) to a set of states represented symbolically~\cite{CousotCousot_JLC92}.
For instance, if we are only interested in scalar numerical program variables, such a set may be a convex polyhedron (the set of solutions of a system of linear inequalities)~\cite{CousotHalbwachs78,Halbwachs_PhD,PPL,BagnaraHZ08SCP}.

In such an analysis, information may flow forward \avirer{(one computes the polyhedron after a program statement as the image by the semantics of the statement, or a super-set thereof, of the polyhedron before)} or backward;
forward program analysis computes super-sets of the states reachable from the initialization of the program, backward program analysis computes super-sets of the states co-reachable from some property of interest (for instance, the violation of an assertion).
In forward analysis, control-flow joins correspond to convex hulls if using convex polyhedra (more generally, they correspond to least upper bounds in a lattice); in backward analysis, it is control-flow splits that correspond to convex hulls.

It is a known limitation of program analysis by abstract interpretation that this convex hull, or more generally, least upper bound operation, may introduce states that cannot occur in the real program: for instance, the convex hull of the intervals $[-2,-1]$ and $[1,2]$ is $[-2,2]$, strictly larger than the union of the two.
Such introduction may prevent proving desired program properties, for instance $\neq 0$. The alternative is to keep the union symbolic (e.g. compute using $[-2,-1] \cup [1,2]$) and thus compute in the \emph{disjunctive completion}
of the lattice, but the number of terms in the union may grow exponentially with the number of successive tests in the program to analyze, not to mention difficulties for designing suitable widening operators for enforcing the convergence of fixpoint iterations~\cite{PPL,BagnaraHZ08SCP,DBLP:journals/sttt/BagnaraHZ07}.
The exponential growth of the number of terms in the union may be controlled by heuristics that judiciously apply least upper bound operations, as in the \emph{trace partitioning domain} \cite{Rival_Mauborgne_TOPLAS07} implemented in the Astr\'ee analyzer~\cite{ASTREE_PLDI03,ASTREE_ESOP05}.

Assuming we are  interested in a loop-free program fragment, the above approach of keeping symbolic unions gives the same results as performing the analysis separately over every path in the fragment.
A recent method for finding disjunctive loop invariants \cite{DBLP:conf/pldi/GulwaniZ10} is based on this idea: each path inside the loop body is considered separately.
Two recent proposals use SMT-solving \cite{Kroening_Strichman_08} as a decision procedure for the satisfiability of first-order arithmetic formulas in order to enumerate only paths that are needed for the progress of the analysis \cite{Gawlitza_Monniaux_ESOP11,Monniaux_Gonnord_SAS11}. They can equivalently be seen as analyses over a multigraph of transitions between some distinguished control nodes. This multigraph has an exponential number of edges, but is never explicitly represented in memory; instead, this graph is \emph{implicitly} or \emph{succinctly} represented: its edges are enumerated as needed as solutions to SMT problems.

An additional claim in favor of the methods that distinguish paths inside the loop body \cite{DBLP:conf/pldi/GulwaniZ10,Monniaux_Gonnord_SAS11} is that they tend to generate better invariants than methods that do not, by behaving better with respect to the \emph{widening operators} \cite{CousotCousot_JLC92}
used for enforcing convergence when searching for loop invariants by Kleene iterations. A related technique, \emph{guided static analysis} \cite{DBLP:conf/sas/GopanR07}, computes successive loop invariants for increasing subsets of the transitions taken into account, until all transitions are considered; again, the claim is that this approach avoids some gross over-approximation introduced by widenings.

All these methods improve the precision of the analysis by keeping the
same abstract domain (say, convex polyhedra) but changing the
operations applied and their ordering. An alternative is to change the
abstract domain (e.g. octagons, convex polyhedra
\cite{DBLP:journals/lisp/Mine06}), or the widening
operator~\cite{BagnaraHRZ05SCP,Polka:FMSD:97}.

\avirer{There are many possible combinations of the above techniques, and it is not
evident which ones perform more or less precisely or more or less efficiently on real-life examples. One needs to experiment.
Unfortunately, the published literature on the subject lacks experimental comparative assessments. One purpose of this article is therefore to propose such experimental results.}

This article makes the following contributions:
\begin{compactenum}
\item We recast the guided static analysis technique from \cite{DBLP:conf/sas/GopanR07} on the expanded multigraph from \cite{Monniaux_Gonnord_SAS11}, considering entire paths instead of individual transitions, using SMT queries and binary decision diagrams\label{contr:guided_multigraph} (See \S\ref{sec:guided_multigraph}).
\item We improve the technique for obtaining disjunctive invariants from \cite{DBLP:conf/pldi/GulwaniZ10} by replacing the explicit exhaustive enumeration of paths by a sequence of SMT queries\label{contr:disjunctive} (See \S\ref{sec:disjunctive}).
\item We implemented these techniques, in addition to ``classical'' iterations and the original guided static analysis, inside a prototype static analyzer.
This tool uses the LLVM bitcode format \cite{Lattner:2004:LCF:977395.977673,LLVM_langref} as input, which can be produced by compilation from C, C++ and Fortran, enabling it to be run on many real-life programs.
It uses the APRON library \cite{DBLP:conf/cav/JeannetM09}, which
supports a variety of abstract domains for numerical variables, from
which we can choose with minimal changes to our analyzer.

\avirer{
Our tool uses a \emph{single static assignment} (SSA) intermediate representation, which allows us to perform cheap symbolic propagations and simplifications,
described in Section~\ref{sec:analysis-algorithm}.
}

\item We conducted extensive experiments with this tool, on real-life programs\avirer{: (i) fixing the abstract domain, varying the iteration technique (\S\ref{sec:compare_techniques}) (ii) fixing the iteration technique, varying the abstract domain (\S\ref{sec:compare_domains})}.
\end{compactenum}

\section{Bases}
\subsection{Static Analysis by Abstract Interpretation}
\label{sec:static_analysis}
\MM{On peut peut-être réduire un peu vu que tout le monde connait à
  SAS ?}

Let $X$ be the set of possible states of the program variables; for instance, if the program has 3 unbounded integer variables, then $X = \ZZ^3$. The set $\parts{X}$ of subsets of $X$, partially ordered by inclusion, is the \emph{concrete domain}. An \emph{abstract domain} is a set $\abstr{X}$ equipped with a partial order $\sqsubseteq$ (the associated strict order being $\sqsubset$); for instance, it can be the domain of convex polyhedra in $\QQ^3$ ordered by geometric inclusion. \avirer{In this article, all abstract domains are supposed to contain machine-representable objects, and all $\sqsubseteq$ order relations are supposed to be decidable.}
The concrete and abstract domains are connected by a monotone \emph{concretization} function $\gamma: \left(\abstr{X},\sqsubseteq\right) \rightarrow (\parts{X},\subseteq)$: an element $\abstr{x} \in \abstr{X}$ represents a set $\gamma(\abstr{x})$.%
%

We also assume a join operator $\sqcup: \abstr{X} \times \abstr{X} \rightarrow \abstr{X}$, with infix notation; in practice, it is generally a least upper bound operation, but we only need it to satisfy $\gamma(\abstr{x}) \cup \gamma(\abstr{y}) \subseteq \gamma(\abstr{x} \sqcup \abstr{y})$ for all $\abstr{x},\abstr{y}$.

Classically, one considers the control-flow graph of the program, with edges labeled with concrete transition relations (e.g. $x' = x+1$ for an instruction \lstinline|x = x+1;|), and attaches an abstract element to each control point.
A concrete transition relation $\tau \subseteq X \times X$ is replaced by an abstract \emph{forward abstract transformer} $\abstr{\tau}: \abstr{X} \rightarrow \abstr{X}$, such that
\(
\forall \abstr{x} \in \abstr{X}, x,x' \in X,~
x \in \gamma(\abstr{x}) \land (x,x') \in \tau \implies
x' \in \gamma \circ \abstr{\tau}(\abstr{x})
\).
It is easy to see that if to any control point $p \in P$ we attach an abstract element $\abstr{x}_p$ such that
(i) for any $p$, $\gamma(\abstr{x}_p)$ includes all initial states possible at control node $p$
(ii) for any $p,p'$, $\abstr{\tau}_{p,p'} (\abstr{x}_p) \sqsubseteq \abstr{x}_{p'}$, noting $\tau_{p,p'}$ the transition from $p$ to~$p'$, then $(\gamma(\abstr{x}_p))_{p \in P}$ form an \emph{inductive invariant}: by induction, when the control point is $p$, the program state always lies in $\gamma(\abstr{x}_p)$.

\emph{Kleene iterations} compute such an inductive invariant as the stationary limit, if it exists, of the following system: for each $p$, initialize $\abstr{x}_p$ such that $\gamma(\abstr{x}_p)$ is a superset of the initial states at point $p$; then iterate the following:
if $\abstr{\tau}_{p,p'} (\abstr{x}_p) \not\sqsubseteq \abstr{x}_{p'}$, replace $\abstr{x}_{p'}$ by $\abstr{x}_{p'} \sqcup \abstr{\tau}_{p,p'} (\abstr{x}_p)$.
Such a stationary limit is bound to exist if $\abstr{X}$ has no infinite ascending chain $a_1 \sqsubset a_2 \sqsubset \dots$; this condition is however not met by domains such as intervals or convex polyhedra.

\emph{Widening-accelerated Kleene iterations} proceed by replacing $\abstr{x}_{p'} \sqcup \abstr{\tau}_{p,p'} (\abstr{x}_p)$ by $\abstr{x}_{p'} \widening (\abstr{x}_{p'} \sqcup \abstr{\tau}_{p,p'} (\abstr{x}_p))$ where $\widening$ is a \emph{widening operator}: for all $\abstr{x},\abstr{y}$, $\gamma(\abstr{y}) \subseteq \gamma(\abstr{x} \widening \abstr{y})$, and any sequence $\abstr{u}_1,\abstr{u}_2,\dots$ of the form $\abstr{u}_{n+1} = \abstr{u}_n \widening \abstr{v}_n$, where $\abstr{v}_n$ is another sequence, become stationary.
The stationary limit $(\abstr{x}_p)_{p \in P}$, defines an inductive invariant $(\gamma(\abstr{x}_p))_{p \in P}$. Note that this invariant is not, in general, the least one expressible in the abstract domain, and may depend on the iteration ordering (the successive choices~$p,p'$).

Once an inductive invariant $\gamma((\abstr{x}_p)_{p \in P})$ has been obtained,
one can attempt \emph{decreasing} or \emph{narrowing} iterations to reduce it.
In their simplest form, this just means running the following operation until a
fixpoint or a maximal number of iterations are reached: for any $p'$, replace
$\abstr{x}_{p'}$ by $\abstr{x}_{p'} \cap \left(\bigsqcup_{p \in P}
\abstr{\tau}_{p,p'} (\abstr{x}_p)\right)$. The result also defines an inductive
invariant. These decreasing iterations are indispensable to recover properties
from guards (tests) in the program in most iteration settings; unfortunately,
certain loops, particularly those involving identity (no-operation) transitions,
may foil them: the iterations immediately reach a fixpoint and do not decrease
further (see example in \S\ref{subsec:rate_lim}). Sections \ref{sec:guided} and
\ref{sec:path_focusing} describe techniques that work around this problem.

\MM{J'avais insisté pour qu'on écrive un truc comme ça, mais si on ne
  met plus l'accent sur le côté expérimental, on peut sans doute
  supprimer ce paragraphe completement ?}
\avirer{
Widening operations have a somewhat counterintuitive behavior: the
result of an analysis (postcondition) is not necessarily monotonic
with respect to the precondition.
This means that, when using widenings, being locally more precise \avirer{(for instance, by using more expressive abstract domains, or widening operators that climb more slowly \cite{BagnaraHRZ05SCP})} does not necessarily translate into higher final precision.
This justifies our recourse to evaluation on real, large programs~(\S~\ref{sec:compare_techniques}).
}

\subsection{SMT-solving}
Boolean satisfiability (SAT) is the canonical NP-complete problem: given a propositional formula (e.g. $(a \lor \neg b) \land (\neg a \lor b \lor \neg c)$), decide whether it is satisfiable --- and, if so, output a satisfying assignment.
Despite an exponential worst-case complexity, the DPLL algorithm \cite{Kroening_Strichman_08,Handbook_SAT} solves many useful SAT problems in practice.

SAT was extended to \emph{satisfiability modulo theory} (SMT): in addition to propositional literals, SMT formulas admit atoms from a theory.
For instance, the theories of linear integer arithmetic (LIA) and linear real arithmetic (LRA) have atoms of the form $a_1 x_1 + \dots + a_n x_n \bowtie C$ where $a_1,\dots,a_n,C$ are integer constants, $x_1,\dots,x_n$ are variables (interpreted over $\ZZ$ for LIA and $\RR$ or $\QQ$ for LRA), and $\bowtie$ is a comparison operator $=,\neq,<,\leq,>,\geq$.
Satisfiability for LIA and LRA is NP-complete, yet tools based on DPLL(T) approach \cite{Kroening_Strichman_08,Handbook_SAT} solve many useful SMT problems in practice. All these tools provide a \emph{satisfying assignment} if the problem is satisfiable.

\avirer{
Most SMT solvers, including Z3,
Yices,
and all those supporting the full SMTLIB2 standard \cite{BarST-SMTLIB},
offer an \emph{incremental} interface: the client program specifies the formula as an initially empty conjunction, to which additional constraints are added, and calls a ``check'' function answering whether it is satisfiable; 
it may then backtrack some of the constraints and add other ones without restarting from scratch.
}

\subsection{A Simple, Motivating Example}
\label{subsec:rate_lim}
Consider the following program, adapted from \cite{Monniaux_Gonnord_SAS11}, where \lstinline|input($a$, $b$)| stands for a nondeterministic input in $[a,b]$ (the control-flow graph on the right depicts the loop body, $s$ is the start node and $e$ the end node):

\begin{minipage}{9cm}
\begin{lstlisting}
void rate_limiter() {
  int x_old = 0;
  while (1) {
    int x = input(-100000, 100000);
    if (x > x_old+10) x = x_old+10;
    if (x < x_old-10) x = x_old-10;
    x_old = x;
} }
\end{lstlisting}
\end{minipage}%
\begin{minipage}[c]{2cm}
\begin{tikzpicture}[->,>=stealth',auto,node distance=.2cm and .5cm,
                    semithick,font=\footnotesize]

	\node[PRstate] (n0) {$s$};
	\node[state] (n0x) [below =of n0] {};
	\node[state] (n1) [below left=of n0x] {};
	\node[state] (n3) [below right=of n1] {};
	\node[state] (n4) [below left=of n3] {};
	\node[state] (n6) [below right=of n4] {};
	\node[PRstate] (n7) [below =of n6] {$e$};

  \path [transition] (n0) -- (n0x);
  \path [transition] (n0x) -- (n1);
  \path [transition] (n0x) edge[bend left] (n3);
  \path [transition] (n1) -- (n3);
  \path [transition] (n3) -- (n4);
  \path [transition] (n3) edge[bend left] (n6);
  \path [transition] (n4) -- (n6);
  \path [transition] (n6) -- (n7);
\end{tikzpicture}
\end{minipage} 

This program implements a construct commonly found in control programs (in e.g.
automotive or avionics): a rate or slope limiter.
\avirer{For the sake of simplicity, we chose it
to be fed on a nondeterministic input clamped between $[-100000,100000]$, but in a real system it would be integrated in a reactive control loop and its input connected to a complex system with unknown output range.}

The expected inductive invariant is $\lstinline|x_old| \in [-100000,100000]$, but classical abstract interpretation using intervals (or octagons or polyhedra) finds $\lstinline|x_old| \in (-\infty,+\infty)$~\cite{ASTREE_ESOP05}.
Let us briefly see why.

Widening iterations converge to $\lstinline|x_old| \in (-\infty,+\infty)$; let us now see why decreasing iterations fail to recover the desired invariant.
The \lstinline|x > x_old+10| test at line~6, if taken, yields $\lstinline|x_old|
\in (-\infty,99990)$; followed by \lstinline|x = x_old+10|, we obtain
$\lstinline|x| \in (-\infty,\allowbreak 100000)$, and the same after union with the no-operation ``else'' branch. Line~7 yields $\lstinline|x| \in (-\infty,+\infty)$.

We could use ``widening up to'' or ``widening with thresholds'', propagating the ``magic values'' $\pm 100000$ associated to \lstinline|x| into \mbox{\lstinline|x_old|,} but these syntactic approaches cannot directly cope with programs for which $\lstinline|x|  \in [-100000,+100000]$ is itself obtained by analysis.
The guided static analysis of \cite{DBLP:conf/sas/GopanR07} does not perform
better, and also obtains $\lstinline|x_old| \in (-\infty,+\infty)$.

In contrast, let us distinguish all four possible execution paths through the tests at lines 6 and~7. The path through both ``else'' branches is infeasible; the program is thus equivalent to a program with 3 paths:

\begin{minipage}[c]{9cm}
\begin{lstlisting}
void rate_limiter() {
  int x_old = 0;
  while (1) {
    int x = input(-100000, 100000);
    if (x > x_old+10) x_old = x_old+10;
    else if (x < x_old-10) x_old = x_old-10;
    else x_old = x;
} }
\end{lstlisting}
\end{minipage}
\begin{minipage}[c]{2cm}
\begin{tikzpicture}[->,>=stealth',auto,node distance=.2cm and .5cm,
                    semithick,font=\footnotesize]

	\node[PRstate] (n0) {$s$};
	\node (n1) [below left=of n0] {};
	\node (n2) [below right=of n0] {};
	\node (n3) [below right=of n1] {};
	\node (n4) [below left=of n3] {};
	\node (n5) [below right=of n3] {};
	\node (n6) [below right=of n4] {};
	\node (n7) [below left=of n6] {};
	\node (n8) [below right=of n6] {};
	\node[PRstate] (n9) [below right=of n7] {$e$};

  \path [transition] 
		(n0) edge  [out=230, in=130]            node {} (n9);
  \path [transition] 
		(n0) edge  [out=255, in=105]            node {} (n9);
  \path [transition,dotted] 
		(n0) edge  [out=-50, in=50]            node {} (n9);
  \path [transition] 
		(n0) edge  [out=-75, in=75]            node {} (n9);
\end{tikzpicture}
\end{minipage}

Classical interval analysis on this program yields $\lstinline|x_old| \in [-100000,100000]$.
We have transformed the program, manually pruning out infeasible paths; yet in general the resulting program could be exponentially larger than the first, even though not all feasible paths are needed to compute the invariant.

Following recent suggestions \cite{Gawlitza_Monniaux_ESOP11,Monniaux_Gonnord_SAS11}, we avoid this space explosion by keeping the second program implicit while simulating its analysis. This means we work on an implicitly represented transition multigraph \avirer{(Fig.~\ref{fig:multigraph})}; it is succinctly represented by the transition graph of the first program.
Our first contribution (\S\ref{sec:guided_multigraph}) is to recast the ``guided analysis'' from \cite{DBLP:conf/sas/GopanR07} on such a succinct representation of the paths in lieu of the individual transitions.
A similar explosion occurs in disjunctive invariant generation, following \cite{DBLP:conf/pldi/GulwaniZ10}; our second contribution (\S\ref{sec:disjunctive}) applies our implicit representation to their method.

\subsection{Guided Static Analysis}
\label{sec:guided}
\begin{figure}
  \begin{minipage}{.3\linewidth}
\begin{lstlisting}[label=lst:gopan_reps]
int x = 0, y = 0;
while (1) {
  if (x <= 50) y++;
  else y--;
  if (y < 0) break;
  x++;
}
\end{lstlisting}
  \end{minipage}
\hfill
\begin{minipage}{.5\linewidth}
\includegraphics[scale=0.7]{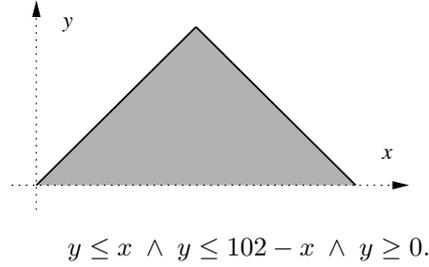}
\begin{equation*}
y \leq x \;\land\; y \leq 102-x \;\land\; y \geq 0.\label{eqn:triangle}
\end{equation*}
\end{minipage}
\caption{Example program and its invariant: the piecewise linear, solid line is the strongest invariant, the grayed polyhedron is its convex hull.}
\label{fig:gopan_reps_invariant} 
\end{figure}

\emph{Guided static analysis} was proposed by \cite{DBLP:conf/sas/GopanR07} as an improvement over classical upward Kleene iterations with widening.
Consider the program in Fig.~\ref{fig:gopan_reps_invariant}, taken from \cite{DBLP:conf/sas/GopanR07}.

Classical iterations on the domain of convex polyhedra \cite{CousotHalbwachs78,BagnaraHRZ05SCP} or octagons \cite{DBLP:journals/lisp/Mine06} start with $x = 0 \land x = 0$, then continue with $x = y \land 0 \leq x \leq 1$.
The widening operator extrapolates from these two iterations and yields $x = y \land x \geq 0$.
From there, the ``else'' branch at line~4 may be taken; with further widening, $0 \leq y \leq x$ is obtained as a loop invariant, and thus the computed loop postcondition is $x \geq 0 \land y = 0$.
Yet the strongest invariant is $(0 \leq x \leq 51 \land y = x) \lor (51 \leq x \leq 102 \land x+y=102)$, and its convex hull, a convex polyhedron (Fig.~\ref{fig:gopan_reps_invariant}).

Intuitively, this disappointing result is obtained because widening extrapolates from the first iterations of the loop, but the loop has two different phases ($x \leq 50$ and $x > 50$) with different behaviors, thus the extrapolation from the first phase is not valid for the second.

Gopan and Reps' idea is to analyze the first phase of the loop with a widening and narrowing sequence, and thus obtain $0 \leq x \leq 50 \land y = x$, and then analyze the second phase, finally obtaining invariant~(\ref{eqn:triangle}); each phase is identified by the tests taken or not taken.

The analysis starts by identifying the tests taken and not taken during the first iteration of the loop, starting in the loop initialization. The branches not taken are pruned from the loop body, yielding:
\begin{lstlisting}[numbers=none]
  while(1) {
    if(x <= 50) y++;
    else break; /* not taken in phase 1 */
    if(y < 0) break;
    x++;
  }
\end{lstlisting}

Analyzing this loop using widening and narrowing on convex polyhedra
or octagons yields the loop invariant $0 \leq x \leq 51 \land y = x$.
Now, the transition at line~4 becomes feasible; and we analyze the
full loop, starting iterations from $0 \leq x \leq 51 \land y = x$,
and obtain invariant~(\ref{eqn:triangle}) in Fig~\ref{fig:gopan_reps_invariant}.

More generally, this analysis method considers an ascending sequence of subsets of the transitions in the loop body \avirer{(left side of Fig.~\ref{fig:multigraph})};
for each subset, an inductive invariant is computed for the program restricted to it.
The starting subset consists in the transitions reachable in one step from the loop initialization.
If for a given subset $S$ in the sequence, no transitions outside $S$ are reachable from the inductive invariant attached to $S$, then iterations stop;
otherwise, add these transitions to $S$ and iterate more.
Termination ensues from the finiteness of the control-flow graph.

\subsection{Path-focusing}
\label{sec:path_focusing}

Monniaux \& Gonnord's \emph{path-focusing} \cite{Monniaux_Gonnord_SAS11} technique
distinguishes the different paths in the program in order to avoid loss of
precision due to merge operations. Since the number of paths may be exponential,
the technique keeps them implicit and computes them when needed using
SMT-solving.
The (accelerated) Kleene iterations (\S\ref{sec:static_analysis}) are computed over a reduced multigraph instead of the classical transition graph.

Let $P$ be the set of control points in the transition graph, 
$P_W \subseteq P$ the set of widening points
such that removing the points in $P_W$ gives an acyclic graph.
One can choose a set $P_R$ such that $P_W \subseteq P_R \subseteq P$.

The set of paths is kept implicit by an SMT formula $\rho$ expressing
the semantics of the program, assuming that the transition semantics can be
expressed within a decidable theory. For an easy construction of $\rho$, 
we also assume that the program is expressed in SSA form, meaning that each
variable is only assigned once in the transition graph. This is not a
restriction, since there exists standard algorithms that transform a program into
an SSA format.

This formula contains Boolean \emph{reachability predicates} $b_i$ for each
control points $p_i \notin P_R$, $b_i^s$ and $b_i^d$ for each $p_i \in P_R$, so
that a path 
$p_{i_1} \rightarrow p_{i_2} \rightarrow \dots \rightarrow p_{i_n}$ 
between two points $p_{i_1}, p_{i_n} \in P_R$ 
can easily be expressed as the
conjunction $b_{i_1}^s \wedge \bigwedge_{2 \leq k < n} b_{i_k} \wedge b_{i_n}^d$.
The Boolean $b_{i}^s$ is $true$ when the path starts at point $p_i$, whereas
$b_i^d$ is $true$ when the path arrives at $p_i$. In other words, we split the
points in $P_R$ into a \emph{source} point, with only outgoing transitions, and
a \emph{destination} point, with only incoming transitions, so that the
resulting graph is acyclic and there are no paths going through control
points in $P_R$.

In order to find focus paths, we solve an SMT formula which is satisfiable when
there exists a path starting at a point $p_i \in P_R$ in a state included in the
current invariant candidate $X_i$, and arriving at a point $p_j \in P_R$ in a
state outside $X_j$. In this case, we construct this path using the model and
update $X_j$. When $p_i = p_j$, meaning that the path is actually a self-loop,
we can apply a widening/narrowing sequence, or even compute the transitive
closure of the loop (or an approximation thereof, or its application to $X_i$)
using abstract acceleration~\cite{DBLP:conf/sas/GonnordH06}.

We assume that we can encode the concrete semantics of the program into the SMT formula, or at least an abstraction thereof at least as precise as the one applied by the abstract interpreter (in simple terms: we want to avoid the case where the SMT solver exhibits a possible path, but the static analyzer realizes that this path is infeasible; this would lead to nontermination, because the SMT solver would exhibit the same path on the next iteration).
A workaround would be to apply \emph{satisfiability modulo path programs} \cite{DBLP:conf/popl/HarrisSIG10}: from each path ruled infeasible by abstract interpretation, extract a blocking clause for the SAT solver underlying the SMT-solver.

\section{Guided Analysis over the Paths}
\label{sec:guided_multigraph}

Guided static analysis, as proposed by \cite{DBLP:conf/sas/GopanR07}, applies to the transition
graph of the program. We now present a new technique applying this analysis on the implicit
multigraph from \cite{Monniaux_Gonnord_SAS11}, thus avoiding control flow merges with
unfeasible paths.
In this section, we use the same notations as \S\ref{sec:path_focusing}.

The combination of these two techniques aims at first discovering a precise
inductive invariant for a subset of paths between two points in $P_R$, 
by the mean of ascending and narrowing iterations. When an
inductive invariant has been found, we add new feasible paths to the subset and
compute an inductive invariant for this new subset, starting with the results
from the previous analysis.
In other words, our technique considers an ascending sequence of
subsets of the paths between two points in $P_R$.
We iterate the operations until the whole program (i.e all the
feasible paths) has been considered. The result will then be an
inductive invariant of the entire program.

The ascending iteration applies path-focusing
\cite{Monniaux_Gonnord_SAS11} to a subset of the multigraph. As
\cite{DBLP:conf/sas/GopanR07}, we do some narrowing, to recover
precision lost by widening, \emph{before}
computing and taking into account new feasible paths. Thus, our
technique combines the advantages of \emph{Guided Static
  Analysis} and \emph{Path-focusing}.


Algorithm~\ref{algo:combined} performs Guided static analysis on the implicitly represented multigraph.
$I_p$ denotes a set of initial states at program point~$p$ (thus $\emptyset$ for most $p$).
The current working subset of paths, noted $P$ and initially empty, is
stored using a compact representation, such as binary decision
diagrams. We also maintain two sets of control points:
\begin{compactitem}
	\item $A'$ : points in $P_R$ that may be the starting points of new
		feasible paths.
	\item $A$ : points in $P_R$ on which we apply the ascending iterations.
	When the abstract value of a control point $p$ is updated, $p$ is
	added to both $A$ and $A'$.
\end{compactitem}

\begin{algorithm}
	\caption{Guided static analysis on implicit multigraph}
	\label{algo:combined}
	\begin{algorithmic}[1] 
	\STATE $A' \gets \{p | P_R / I_p \neq \emptyset\}$
\STATE $A \gets \emptyset$
\STATE $P \gets \emptyset$ \COMMENT{Paths in the current subset}
\FORALL{$p_i \in P_R$}
	\STATE $X_i \gets I_{p_i}$ \label{alg=X-init}
\ENDFOR 
\WHILE{$A' \neq \emptyset$}

\WHILE{$A' \neq \emptyset$} \label{alg=start-add-paths} 
	\STATE Select $p_i \in A'$
	\STATE $A' \gets A' \setminus \{p_i\}$
	\STATE ComputeNewPaths($p_i$) \COMMENT{Update $A$, $A'$ and $P$}\label{alg=computeNewPaths}
\ENDWHILE \label{alg=end-add-paths}

\STATE \COMMENT{ascending iterations on $P$}
\WHILE{$A \neq \emptyset$} \label{alg=start-ascending} 
	\STATE Select $p_i \in A$
	\STATE $A \gets A \setminus \{p_i\}$
	\STATE PathFocusing($p_i$) \COMMENT{Update $A$ and $A'$\label{alg=pf}}
	
\ENDWHILE \label{alg=end-ascending}
\STATE Narrow \label{alg=narrowing}
\ENDWHILE

\RETURN $\{X_i,\ i \in P_R\}$

	\end{algorithmic}
\end{algorithm}

We distinguish three phases in the main loop of the analysis:
\begin{compactenum}
\item \label{step:addingpaths} We start finding a new relevant subset
  $P$ of the graph.
  Either the previous iteration or the initialization led us to a
  state where there are no more paths in the previous subset $P$,
  starting at $p_i$, that make the abstract values of the successors
  grow (otherwise, the SMT solver would not have answered
  ``\emph{unsat}''). Narrowing iterations preserve this property.
  However, there may exist such paths in the entire multigraph, that
  are not in $P$. This phase computes these paths and adds them to
  the subset. This phase is described in~\ref{subsec:addingpaths}
  and corresponds to lines in~\ref{alg=start-add-paths} to
  \ref{alg=end-add-paths} in Algorithm~\ref{algo:combined}.
\item \label{step:ascending} Given a new subset $P$, we search for paths starting at point
  $p_i \in P_R$, such that these paths are in $P$, i.e are included in
  the working subgraph. Each time we find a path, we update the
  abstract value of the destination point of the path. This is the
  phase explained in~\ref{subsec:ascending}, and corresponds to
  lines~\ref{alg=start-ascending} to~\ref{alg=end-ascending} in
  Algorithm~\ref{algo:combined}.
\item  \label{step:narrowing} We perform narrowing iterations the usual way
  (line~\ref{alg=narrowing} in algorithm~\ref{algo:combined}) and
  reiterate from step 1 unless there are no more points to explore,
  i.e. $A' = \emptyset$.
\end{compactenum}

The order of steps is important: narrowing has to be performed before
adding new paths, or spurious new paths would be added to $P$.
Starting with the addition of new paths avoids
doing the ascending iterations on an empty graph.

\subsection{Ascending Iterations by Path-focusing}
\label{subsec:ascending}


For computing an inductive invariant over a subgraph, we use the
Path-focusing algorithm from \cite{Monniaux_Gonnord_SAS11} with special
treatment for self loops (line~\ref{alg=pf} in algorithm~\ref{algo:combined}).

In order to find which path to focus on, we construct an SMT formula $f(p_i)$, whose
model when satisfiable is a path that starts in $p_i$, goes to a successor $p_j
\in P_R$ of $p_i$, such that the image of $X_{i}$ by the path transformation
is not included in the current $X_{j}$.
Intuitively, such a path makes the abstract value $X_{j}$ grow, and thus is
an interesting path to focus on. We loop until the formula becomes unsatisfiable,
meaning that the analysis of $p_i$ is finished.

If we note $Succ(i)$ the set of indices $j$ such that $p_j \in P_R$ is a
successor of $p_i$ in the expanded multigraph, and $X_i$ the abstract value
associated to $p_i$ :
\[
f(p_i) = \rho \wedge b_i^s \wedge 
\displaystyle\bigwedge_{j \in P_R \atop j \neq i} \neg
b_j^s \wedge X_i \wedge \displaystyle\bigvee_{j \in Succ(i)} (b_j^d \wedge
 \neg X_j)
\]
The difference with~\cite{Monniaux_Gonnord_SAS11} is that we do not
work on the entire transition graph but on a subset of it. Therefore we
conjoin the formula $f(p_i)$ with the actual set of working paths,
noted $P$, expressed as a Boolean formula, where the Boolean variables are the
\emph{reachability predicates} of the control points. We can easily construct
this formula from the binary decision diagram using dynamic programming, and
avoiding an exponentially sized formula. In other words, we force the SMT solver
to give us a path included in $P$.
Each time the invariant candidate of a point $p_j$ has been updated, $p_j$ is
inserted into $A'$ since it may be the start of a new feasible paths.

\subsection{Adding New Paths}
\label{subsec:addingpaths}

Our technique computes the fixpoint iterations on an ascending sequence of
subgraphs, until the complete graph is reached.
When the analysis of a subgraph is finished, meaning that the abstract values
for each control point has converged to an inductive invariant for this subgraph,
the next subgraph to work on has to be computed.

This new subgraph contains all the paths from the previous one, and also new
paths that become feasible regarding the current abstract values.
The new paths in $P$ are computed one after another, until no more path
can make the invariant grow. This is line~\ref{alg=computeNewPaths} in
Algorithm~\ref{algo:combined}, which corresponds to
Algorithm~\ref{algo:computepaths}. We also use SMT solving to discover
these new paths, but we subtly change the SMT formula given to the
SMT solver: we now try to find a path that is not yet in $P$, but is feasible
and makes the invariant candidate of its destination grow.
We thus check the satisfiability of the formula
$f'(p_i)$, where:
\[
f'(p_i) = f(p_i)\wedge \neg P
\]
$X_j$ is updated using an abstract union when the point $p_j$ is the target of a new path. This way, further SMT queries do not
compute other paths with the same source and destination if it is not needed
(because these new paths would not make $X_j$ grow, hence would not be returned
by the SMT solver).

\begin{algorithm}
	\caption{ComputeNewPaths}
	\label{algo:computepaths}
	\begin{algorithmic}[1] 
	\WHILE{true}
	\STATE $res \gets SmtSolve\left[
	 f'(p_i) 
	\right]$
	\IF {$res = unsat$}
		\STATE \textbf{break}
	\ENDIF
	\STATE Compute the path $e$ from the model
	\STATE $X_j \gets X_j \sqcup \tau_e(X_i)$
	\STATE $P \gets P \cup \{e\}$
	\STATE $A \gets A \cup \{p_i\}$
	\STATE $A' \gets A' \cup \{p_i\}$
\ENDWHILE

	\end{algorithmic}
\end{algorithm}

When a new path has been found, it is immediately added into $P$. 
We then have to add $p_i$ and $p_j$ into $A$ (since we do not apply widening in
this section) and $p_j$ into $A'$, since $p_j$ may be the starting point of a
new feasible path.

\subsection{Termination}
Termination of this algorithm is guaranteed, because:
\begin{inparaenum}
\item 
the subset of paths $P$ strictly increases at each loop iteration,
and is bounded by the finite set of
paths in the entire graph. 
\item
	when computing new paths, we cunjunct our formula with $\neg P$, meaning
	that we obtain each possible path only once. The number of path is finite,
	so this computation always terminates.
\item
	the Path-focusing iterations terminate because of the properties of
	widening.
\end{inparaenum}

\subsection{Example}

We revise the rate limiter described in \ref{subsec:rate_lim}. In this example,
\emph{Path-focusing} works well because all the paths starting at the
loop header are actually self loops. In such a case, the technique performs a
widening/narrowing sequence or accelerates the loop, thus leading to a precise
invariant. However, in some cases, there also exists paths that are not
self loops, in which case \emph{Path-focusing} applies widening. 
This widening may induce unrecoverable loss of precision.

Suppose the main loop of the rate limiter contains a nested loop like:
\begin{lstlisting}
void rate_limiter() {
  int x_old = 0;
  while (1) {
    int x = input(-100000, 100000);
    if (x > x_old+10) x = x_old+10;
    if (x < x_old-10) x = x_old-10;
    x_old = x;
    while (wait()) {}
} }
\end{lstlisting}

We choose $P_R$ as the set of loop headers of the function, plus the initial
state. In this case, we have three elements in $P_R$.

The main loop in the expanded multigraph has then 4 distinct paths going to the header of the nested loop.

Guided static analysis from \cite{DBLP:conf/sas/GopanR07} yields, at line 3,
$\lstinline|x_old| \in (-\infty,+\infty)$.
Path-focusing \cite{Monniaux_Gonnord_SAS11} also
finds $\lstinline|x_old| \in (-\infty,+\infty)$.
Now, let us see how our technique performs on this example.

\begin{figure}
\begin{minipage}[c]{.50\textwidth}
\begin{tikzpicture}[->,>=stealth',auto,node distance=2.1cm,
                    semithick,font=\footnotesize]

	\node[PRstate] (n00) {$p_1$};
	\node[PRstate] (n0) [below of=n00, yshift=0.8cm] {$p_3$};
	\node[PRstate] (n1) [below of=n0] {$p_{8}$};
	\node (label) [left of=n00, node distance=1cm] {\bf Step 1};

\end{tikzpicture}
\end{minipage}\hfill
\begin{minipage}[c]{.50\textwidth}
\begin{tikzpicture}[->,>=stealth',auto,node distance=2.1cm,
                    semithick,font=\footnotesize]

	\node[PRstate] (n00) {$p_1$};
	\node[PRstate] (n0) [below of=n00, yshift=0.8cm] {$p_3$};
	\node[PRstate] (n1) [below of=n0] {$p_{8}$};
	\node (label) [left of=n00, node distance=1cm] {\bf Step 2};

  \path [transition] 
		(n00) edge  node {$\lstinline|x_old| \gets 0$} (n0);
  \path [transition] 
		(n0) edge [bend left] node [right, xshift=0cm] {
		$\begin{array}{l}
			-10000 \leq \lstinline|x| \leq 10000 \\
			\lstinline|x_old|-10 \leq \lstinline|x|\\
			\lstinline|x| \leq \lstinline|x_old|+10 /\\
			\lstinline|x_old| \gets \lstinline|x|
		\end{array}$
		} (n1);
  \path [transition] 
		(n1) edge [bend left]  node {} (n0);
\end{tikzpicture}
\end{minipage}
\caption{Ascending sequence of subgraphs}
\label{fig:example-graph}
\end{figure}
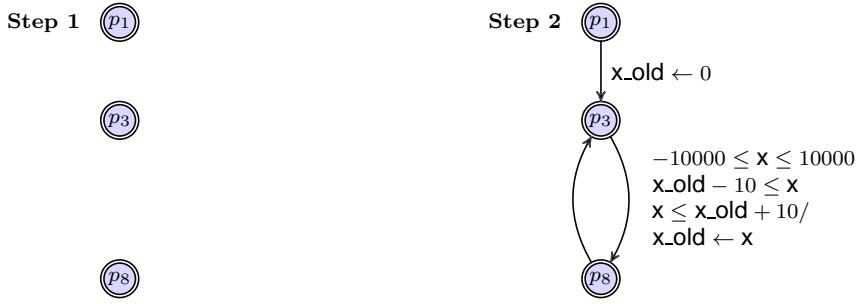

Figure \ref{fig:example-graph} shows the sequence of subset of paths during the
analysis. The points in $P_R$ are noted $p_i$, where $i$ is the corresponding
line in the code: for instance, $p_3$ corresponds to the header of the main
loop.

\begin{compactenum}
	\item The starting subgraph is depicted on Figure
		\ref{fig:example-graph} Step 1. At the beginning, this graph has no
		transitions.
	\item
		We compute the new feasible paths that have to be added into the
		subgraph. We first find the path from $p_1$ to $p_3$ and obtain at $p_3$
		$\lstinline|x_old| = 0$.
		
		The image of $\lstinline|x_old| = 0$ by the path that goes from $p_3$ to
		$p_8$, and that goes through the \emph{else} branch of each 
		\emph{if-then-else}, is $-10 \leq \lstinline|x_old| \leq 10$. This path
		is then added to our subgraph. 
		
		Moreover, there is no other path starting at $p_3$ whose
		image is not in $-10 \leq \lstinline|x_old| \leq 10$.

		Finally, since the abstract value associated to $p_8$ is  
		$-10 \leq \lstinline|x_old| \leq 10$, the path from
                $p_8$ to $p_3$ is
		feasible and is added into $P$. The final subgraph is depicted on Figure
		\ref{fig:example-graph} Step 2.

	\item
		We then compute the ascending iterations by path-focusing. At the end of
		these iterations, we obtain 
$-\infty \leq \lstinline|x_old| \leq +\infty$ for both $p_3$ and $p_8$.
	\item
		We now can apply narrowing iterations, and recover the precision lost by
		widening: we obtain 
		$-10000 \leq \lstinline|x_old| \leq 10000$ at points $p_3$ and $p_8$.
	\item Finally, we compute the next subgraph. The SMT-solver does not find any
		new path that makes the abstract values grow, and the algorithm
		terminates.

\end{compactenum}

Our technique gives us the expected invariant 
$\lstinline|x_old| \in~[-10000,\allowbreak 10000]$. 
Here, only 3 paths out of the 6 have been computed during the analysis. In
practice, depending on the order the SMT-solver returns the paths, other
feasible paths could have been added during the analysis.

In this example, we see that our technique actually combines best of
\emph{Guided Static Analysis} and \emph{Path Focusing}. 

\section{Disjunctive Invariants}
\label{sec:disjunctive}
While many (most?) useful program invariants on numerical variables can be expressed as conjunctions of inequalities and congruences, it is sometimes necessary to introduce disjunctions.
For instance, the loop \lstinline|for (int i=0; i<n; i++) {...}| has head invariant $0 \leq i \leq n \lor (i=0 \land n<0)$.
For this very simple example, a simple syntactic transformation of the control structure (into \lstinline|i=0; if (i<n) do {...} while (i<n)|) is sufficient, but in more complex cases more advanced analyses are necessary
\cite{DBLP:conf/emsoft/BalakrishnanSIG09,DBLP:journals/fmsd/Jeannet03,DBLP:conf/cav/SharmaDDA11,Monniaux_Bodin_APLAS11};
in intuitive terms, they discover \emph{phases} or \emph{modes} in loops.

Gulwani \& Zuleger \cite{DBLP:conf/pldi/GulwaniZ10} proposed a technique for computing disjunctive invariants, by
distinguishing all the paths inside a loop. In
this section, we propose to improve this technique by using SMT queries to find
interesting paths, the objective being to avoid an explicit exhaustive
enumeration of an exponential number of paths.

For each control point $p_i$, we compute a disjunctive invariant
$\bigvee_{1\leq j \leq m_i} X_{i,j}$. We denote by $n_i$ the number of
distinct paths starting at $p_i$.
To perform the analysis, one chooses an integer $\delta_i \in [1,m_i]$, and
a mapping function $\sigma_i: [1,m_i] \times [1,n_i] \mapsto [1,m_i]$.
  The $k$-th path starting fom $p_i$ is denoted $\tau_{i,k}$.
  The image of the $j$-th disjunct $X_{i,j}$ by the path $\tau_{i,k}$ is then
  joined with $X_{i,\sigma(j,k)}$.
Initially, the $\delta_i$-th abstract value contains the initial states of
$p_i$, and all other abstract values contain~$\emptyset$.

For each control point $p_i \in P_R$, $m_i$, $\delta_i$ and $\sigma_i$ can be defined heuristically.
For instance, one could define $\sigma_i$ so that $\sigma_i(j,k)$ only depends on the
last transition of the path, or else construct it dynamically during the
analysis.

Our method improves this technique in two ways :
\begin{compactitem}
\item Instead of enumerating the whole set of paths, we keep them implicit and
compute them only when needed.

\item At each loop iteration of the original algorithm \cite{DBLP:conf/pldi/GulwaniZ10}, an image by each path inside the loop is computed for each disjunct of the invariant candidate.
Yet, many of these images may be redundant: for instance, if our invariant candidate is $(0 \leq x \leq 10 \land 0 \leq y \leq 1000) \lor (x < -10 \land y < -10)$, then there is no point enumerating paths whose image is included in this invariant candidate.
In our approach, we compute such an image only if it makes the resulting abstract value grow.
\end{compactitem}

Our improvement consists in a modification of the SMT formula we solve in
\ref{sec:guided_multigraph}.
We introduce in this formula Boolean variables $\{d_j, 1 \leq j \leq m\}$, so
that we can easily find in the model which abstract value of the
disjunction of the source point has
to be chosen to make the invariant of the destination grow.
The resulting formula that is given to the SMT solver is defined
by $g(p_i)$.
When the formula is satisfiable, we know that the index $j$ of the starting
disjunct that has to be chosen is the one for which the associate Boolean value
$d_j$ is \emph{true} in the model. Then, we can easily compute the value of 
$\sigma_i(j,k)$, thus know the index of the disjunct to join with.
\begin{equation*}
g(p_i) = \rho \wedge b_i^s \wedge 
\displaystyle\bigwedge_{j \in P_R \atop j \neq i} \neg b_j^s 
 \wedge 
\displaystyle\bigvee_{1 \leq k \leq m_i}\!\! (d_k \wedge X_{i,k} \wedge \bigwedge_{l \neq k}
\neg d_l)
\wedge
\displaystyle\bigvee_{j \in Succ(i)}\!\!
(b_j^d \wedge \bigwedge_{1 \leq k \leq m_i}\!\! (\neg X_{j,k}))
\end{equation*}

In our algorithm, the initialization of the abstract values slightly differs from
algorithm \ref{algo:combined} line~\ref{alg=X-init}, since we now have to
initialize each disjunct. Instead of Line~\ref{alg=X-init}, we initialize $X_{i,k}$ with $\perp$ for all
$k \in \{1,..,m_i\} \setminus \{\delta_i\}$, and $X_{i,\delta_i}$ with
$\gets I_{p_i}$.


Furthermore, the Path-focused algorithm (line~\ref{alg=pf} from algorithm
\ref{algo:combined}) is enhanced to deal with disjunctive invariants, and
is detailed in algorithm \ref{algo:disjunctive}.

The \emph{Update} function can classically assign  $X_{i,\sigma_i(j,k)} \widening (X_{i,\sigma_i(j,k)} \sqcup
\tau_{i,k}(X_{i,j}))$ to $X_{i,\sigma_i(j,k)}$, or can integrate the special treatment for self loops
proposed by \cite{Monniaux_Gonnord_SAS11}, with widening/narrowing sequence or
acceleration.

\begin{algorithm}[!htb]
\caption{Disjunctive invariant computation with implicit paths}\label{gulwani2}
\label{algo:disjunctive}
\begin{algorithmic}[1] 
\WHILE {true}
	\STATE $res \gets SmtSolve\left[g(p_i)\right]$
	\IF {$res = unsat$}
		\STATE \textbf{break}
	\ENDIF
	\STATE Compute the path $\tau_{i,k}$ from $res$ 
	\STATE Take $j \in \{ l | d_l = true\}$ 
	\STATE Update($X_{i,\sigma_i(j,k)})$
\ENDWHILE
\end{algorithmic}
\end{algorithm}

We experimented with a heuristic
of dynamic construction of the $\sigma_i$ functions, adapted from
\cite{DBLP:conf/pldi/GulwaniZ10}. 
For each control point $p_i \in P_R$, we start with one single disjunct
($m_i = 1$) and define $\delta_i = 1$.
$M$ denotes an upper bound on the number of disjuncts per control point.

The $\sigma_i$ functions take as parameters the index of the starting abstract
value, and the path we focus on. 
Since we dynamically construct these functions
during the analysis, we store their already computed image into a
compact representation, such as Algebraic Decision Diagrams.
$\sigma_i(j,k)$ is then constructed on the fly only when needed, and computed
only once.
When the value of $\sigma_i(j,k)$ is required but undefined, we first compute
the image of the
abstract value $X_{i,j}$ by the path indexed by $k$, and try to find an existing
disjunct of index $j'$ so that the least upper bound of the two abstract values is exactly their union (using SMT-solving).
If such an index exists, then we set $\sigma_i(j,k) = j'$.
Otherwise:
\begin{compactitem}
	\item if $m_i < M$, we increase $m_i$ by $1$ and define $\sigma_i(j,k) = m_i$
	\item if $m_i = M$, we define $\sigma_i(j,k) = M$ 
\end{compactitem}
The main difference with the original
algorithm~\cite{DBLP:conf/pldi/GulwaniZ10} is that we construct
$\sigma_i(j,k)$ using SMT queries instead of enumerating a possibly
exponential number of paths to find a solution.

\section{Implementation and Experimental Comparisons}
\label{sec:experiments}

We have implemented our proposed solutions inside a prototype of intraprocedural
static analyzer called PAGAI,
as well as the classical abstract interpretation algorithm, and the state-of-the-art
techniques \emph{Path Focusing} \cite{Monniaux_Gonnord_SAS11} and \emph{Guided
Static Analysis} \cite{DBLP:conf/sas/GopanR07}. It is
available online at \url{https://forge.imag.fr/projects/pagai/}. The
implementation is documented in~\cite{julien-henry-m2r}.
\MM{pas idéal de citer un rapport de M2, mais ça a des chances de
  faire taire les reviewers qui voudraient des détails sur comment
  c'est fait, et on n'a pas la place de mettre plus. Le
  paragraphe~\ref{sec:analysis-algorithm} a été supprimé depuis
  PLDI ...}

%
PAGAI operates over LLVM bitcode \cite{LLVM_langref,Lattner:2004:LCF:977395.977673}, which is a target for several compilers, most notably Clang (supporting C and C++) and llvm-gcc (supporting C, C++, Fortran and Ada).
Abstract domains are provided by the APRON library \cite{DBLP:conf/cav/JeannetM09}, and include convex polyhedra (from the builtin Polka ``PK'' library), octagons, intervals, and linear congruences.
For SMT-solving, our analyzer uses Yices
\cite{DBLP:conf/cav/DutertreM06} or Microsoft Z3~\cite{DBLP:conf/tacas/MouraB08}.

PAGAI currently neither models the memory heap nor performs interprocedural analysis. Instead, LLVM optimization phases are applied prior to analysis, in order to inline non-recursive function calls and lift certain memory accesses to operations on explicit numerical variables
(e.g. \lstinline|y=t[0]*t[0];| preceded by \lstinline|t[0]=x;| without any aliased write in between is replaced by \lstinline|y=x*x;|).
The remaining memory reads are considered as indeterminates, and memory writes are ignored; this is a sound abstraction.

\begin{figure}[!htb]%
      \centering%
\begingroup
  \makeatletter
  \providecommand\color[2][]{%
    \GenericError{(gnuplot) \space\space\space\@spaces}{%
      Package color not loaded in conjunction with
      terminal option `colourtext'%
    }{See the gnuplot documentation for explanation.%
    }{Either use 'blacktext' in gnuplot or load the package
      color.sty in LaTeX.}%
    \renewcommand\color[2][]{}%
  }%
  \providecommand\includegraphics[2][]{%
    \GenericError{(gnuplot) \space\space\space\@spaces}{%
      Package graphicx or graphics not loaded%
    }{See the gnuplot documentation for explanation.%
    }{The gnuplot epslatex terminal needs graphicx.sty or graphics.sty.}%
    \renewcommand\includegraphics[2][]{}%
  }%
  \providecommand\rotatebox[2]{#2}%
  \@ifundefined{ifGPcolor}{%
    \newif\ifGPcolor
    \GPcolorfalse
  }{}%
  \@ifundefined{ifGPblacktext}{%
    \newif\ifGPblacktext
    \GPblacktexttrue
  }{}%
  \let\gplgaddtomacro\g@addto@macro
  \gdef\gplbacktext{}%
  \gdef\gplfronttext{}%
  \makeatother
  \ifGPblacktext
    \def\colorrgb#1{}%
    \def\colorgray#1{}%
  \else
    \ifGPcolor
      \def\colorrgb#1{\color[rgb]{#1}}%
      \def\colorgray#1{\color[gray]{#1}}%
      \expandafter\def\csname LTw\endcsname{\color{white}}%
      \expandafter\def\csname LTb\endcsname{\color{black}}%
      \expandafter\def\csname LTa\endcsname{\color{black}}%
      \expandafter\def\csname LT0\endcsname{\color[rgb]{1,0,0}}%
      \expandafter\def\csname LT1\endcsname{\color[rgb]{0,1,0}}%
      \expandafter\def\csname LT2\endcsname{\color[rgb]{0,0,1}}%
      \expandafter\def\csname LT3\endcsname{\color[rgb]{1,0,1}}%
      \expandafter\def\csname LT4\endcsname{\color[rgb]{0,1,1}}%
      \expandafter\def\csname LT5\endcsname{\color[rgb]{1,1,0}}%
      \expandafter\def\csname LT6\endcsname{\color[rgb]{0,0,0}}%
      \expandafter\def\csname LT7\endcsname{\color[rgb]{1,0.3,0}}%
      \expandafter\def\csname LT8\endcsname{\color[rgb]{0.5,0.5,0.5}}%
    \else
      \def\colorrgb#1{\color{black}}%
      \def\colorgray#1{\color[gray]{#1}}%
      \expandafter\def\csname LTw\endcsname{\color{white}}%
      \expandafter\def\csname LTb\endcsname{\color{black}}%
      \expandafter\def\csname LTa\endcsname{\color{black}}%
      \expandafter\def\csname LT0\endcsname{\color{black}}%
      \expandafter\def\csname LT1\endcsname{\color{black}}%
      \expandafter\def\csname LT2\endcsname{\color{black}}%
      \expandafter\def\csname LT3\endcsname{\color{black}}%
      \expandafter\def\csname LT4\endcsname{\color{black}}%
      \expandafter\def\csname LT5\endcsname{\color{black}}%
      \expandafter\def\csname LT6\endcsname{\color{black}}%
      \expandafter\def\csname LT7\endcsname{\color{black}}%
      \expandafter\def\csname LT8\endcsname{\color{black}}%
    \fi
  \fi
  \setlength{\unitlength}{0.0500bp}%
  \begin{picture}(5760.00,3528.00)%
    \gplgaddtomacro\gplbacktext{%
      \csname LTb\endcsname%
      \put(594,966){\makebox(0,0)[r]{\strut{} 0}}%
      \put(594,1253){\makebox(0,0)[r]{\strut{} 2}}%
      \put(594,1540){\makebox(0,0)[r]{\strut{} 4}}%
      \put(594,1827){\makebox(0,0)[r]{\strut{} 6}}%
      \put(594,2115){\makebox(0,0)[r]{\strut{} 8}}%
      \put(594,2402){\makebox(0,0)[r]{\strut{} 10}}%
      \put(594,2689){\makebox(0,0)[r]{\strut{} 12}}%
      \put(594,2976){\makebox(0,0)[r]{\strut{} 14}}%
      \put(594,3263){\makebox(0,0)[r]{\strut{} 16}}%
      \put(1306,834){\rotatebox{-45}{\makebox(0,0)[l]{\strut{}G/S}}}%
      \put(1885,834){\rotatebox{-45}{\makebox(0,0)[l]{\strut{}PF/S}}}%
      \put(2465,834){\rotatebox{-45}{\makebox(0,0)[l]{\strut{}PF/G}}}%
      \put(3045,834){\rotatebox{-45}{\makebox(0,0)[l]{\strut{}G+PF/PF}}}%
      \put(3624,834){\rotatebox{-45}{\makebox(0,0)[l]{\strut{}G+PF/G}}}%
      \put(4204,834){\rotatebox{-45}{\makebox(0,0)[l]{\strut{}G+PF/S}}}%
      \put(4783,834){\rotatebox{-45}{\makebox(0,0)[l]{\strut{}DIS/G+PF}}}%
      \put(146,2115){\rotatebox{-270}{\makebox(0,0){\strut{}percentage of control points}}}%
    }%
    \gplgaddtomacro\gplfronttext{%
      \put(2442,3090){\makebox(0,0)[r]{\strut{}$\subsetneq$}}%
      \put(2442,2870){\makebox(0,0)[r]{\strut{}$\supsetneq$}}%
      \put(2442,2650){\makebox(0,0)[r]{\strut{}uncomparable}}%
    }%
    \gplbacktext
    \put(0,0){\includegraphics{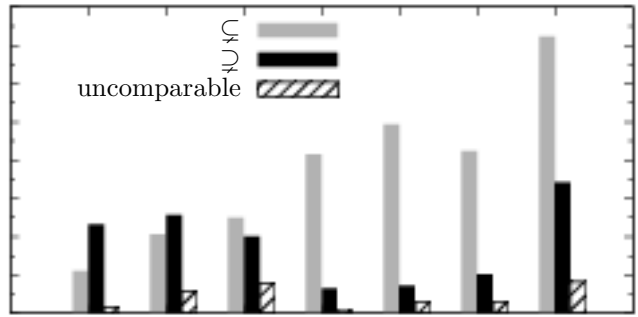}}%
    \gplfronttext
  \end{picture}%
\endgroup

  \caption{Comparison of the abstract values obtained on several open-source projects. The table shows
  their respective number of lines of code, number of control points in $P_R$, and execution time
  on various techniques.
  Techniques are classical abstract
  interpretation (S), \emph{Guided Static
  Analysis} (G), \emph{Path-focused} technique (PF), our combined technique
  (G+PF), and its version with disjunctive invariants (DIS).
  The $\subsetneq$ bars (resp. $\supsetneq$) gives the percentage of invariants stronger (more precise; smaller with respect to inclusion) with the left-side (resp. right-side) technique,
and ``uncomparable'' gives the percentage of invariants that are uncomparable, i.e
neither greater nor smaller;
the code points where both invariants are equal make up the remaining percentage.}
  \label{fig:techniques}
  \label{fig:projects}
\end {figure}

We conducted extensive experiments on real-life programs in order to compare the
different techniques, mostly on open-source projects (Fig.~\ref{fig:projects}) written in C, C++ and Fortran.
These results confirm that our combined technique improve the
analysis in comparison with the two techniques taken individually, at
a reasonable cost. The extension with disjunctive invariants increases
precision in many cases, but with higher cost in terms of execution
time.

\MM{TODO: résumer les conclusions sur les résultats}

\section{Conclusion and Future Prospects}
Roughly, an analysis by abstract interpretation is defined by the choice of an iteration strategy and an abstract domain. In this article, we demonstrated that changes in the iteration algorithm can significantly improve precision, sometimes while improving analysis times.

A common criticism of analysis techniques based on SMT-solving is that they
do not scale up. Yet, our experiments show that, for numerical properties,
they scale up to the size of typical functions and loops.
It is however
quite certain that, naively applied, they cannot scale to the kind of
programs targeted by e.g. the Astr\'ee tool, that is, a dozens or hundreds
of thousands of lines of code in a single loop operating over similar numbers
of remanent variables.
Actually, for such applications, only (quasi-)linear
algorithms scale up, and ``cheap'' abstract domains such as octagons ($O(n^3)$ where $n$ is the number of variables) are not applied to the full variable set, but to restricted subsets thereof.
It thus seems reasonable that techniques such as considering ``packs'' of related variables, slicing, etc. may similarly help SMT-based techniques to scale to global analyses.

\begin{table}[!htb]\centering
      \begin{tabular}{|l|r|r|r|r|r|r|r|} \hline
        \multicolumn{1}{|c|}{} &
        \multicolumn{2}{c|}{Size} &
        \multicolumn{5}{c|}{Execution time (seconds)} \\ \hline
        \multicolumn{1}{|c|}{Name} & 
        \multicolumn{1}{c|}{kLOC}  & 
        \multicolumn{1}{c|}{$|P_R|$}  &
        \multicolumn{1}{c|}{\textbf{S}} &
        \multicolumn{1}{c|}{\textbf{G}} &
        \multicolumn{1}{c|}{\textbf{PF}} &
        \multicolumn{1}{c|}{\textbf{G+PF}} &
	\multicolumn{1}{c|}{\textbf{DIS}} \\ 
        \hline
		a2ps-4.14       &  55   &  2012   & 23 & 74 & 34 & 115 & 162 \\    
gawk-4.0.0      &  59   &  902    & 15 & 46 & 12 & 40 & 50 \\      
gnuchess-6.0.0  &  38   &  1222   & 50 & 220 & 81 & 312 & 351 \\   
gnugo-3.8       &  83   &  2801   & 77 & 159 & 92 & 766 & 1493 \\  
grep-2.9        &  35   &  820    & 41 & 85 & 22 & 65 & 122 \\     
gzip-1.4        &  27   &  494    & 22 & 268 & 91 & 303 & 230 \\   
lapack-3.3.1    &  954  &  16422  & 294 & 3740 & 3773 & 8159 & 10351 \\
make-3.82       &  34   &  993    & 67 & 108 & 53 & 109 & 257 \\   
tar-1.26        &  73   &  1712   & 37 & 218 & 115 & 253 & 396 \\  

	\hline
      \end{tabular}
\caption{Execution times for various techniques}
\label{tab:time}
\end{table}

\MM{Bon, c'est vraiment frustrant de se limiter à ça, mais pour écrire
  plus, faut trouver de la place}

We compared the precision of different techniques and abstract domains by comparing the invariants for the inclusion ordering. A better metric is perhaps to take a client analysis --- such as the detection of overflows and array bound violations --- and compare the rates of alarms.

We focused on numerical properties, because they are supported by easily available abstract libraries. Yet, in most programs, properties of data structures are important for proving interesting properties. Further investigations are needed not only on good abstractions for pointers (many are already known) but also on their conversion to SMT problems.


\FloatBarrier

\bibliographystyle{dmplainnat}
\bibliography{implicitization}

\begin{thebibliography}{30}
\providecommand{\natexlab}[1]{#1}
\providecommand{\url}[1]{\texttt{#1}}
\expandafter\ifx\csname urlstyle\endcsname\relax
  \providecommand{\doi}[1]{doi: #1}\else
  \providecommand{\doi}{doi: \begingroup \urlstyle{rm}\Url}\fi

\bibitem[Bagnara et~al.()Bagnara, Hill, and Zaffanella]{PPL}
Roberto Bagnara, Patricia~M. Hill, and Enea Zaffanella.
\newblock \emph{The Parma Polyhedra Library, version 0.9}.
\newblock URL \url{http://www.cs.unipr.it/ppl}.

\bibitem[Bagnara et~al.(2005)Bagnara, Hill, Ricci, and
  Zaffanella]{BagnaraHRZ05SCP}
Roberto Bagnara, Patricia~M. Hill, Elisa Ricci, and Enea Zaffanella.
\newblock Precise widening operators for convex polyhedra.
\newblock \emph{Science of Computer Programming}, 58\penalty0 (1--2):\penalty0
  28--56, October 2005.
\newblock \doi{10.1016/j.scico.2005.02.003}.

\bibitem[Bagnara et~al.(2006)Bagnara, Hill, and
  Zaffanella]{DBLP:journals/sttt/BagnaraHZ07}
Roberto Bagnara, Patricia~M. Hill, and Enea Zaffanella.
\newblock Widening operators for powerset domains.
\newblock \emph{International Journal on Software Tools for Technology Transfer
  (STTT)}, 8\penalty0 (4-5):\penalty0 449--466, August 2006.
\newblock \doi{10.1007/s10009-005-0215-8}.

\bibitem[Bagnara et~al.(2008)Bagnara, Hill, and Zaffanella]{BagnaraHZ08SCP}
Roberto Bagnara, Patricia~M. Hill, and Enea Zaffanella.
\newblock The {Parma Polyhedra Library}: Toward a complete set of numerical
  abstractions for the analysis and verification of hardware and software
  systems.
\newblock \emph{Science of Computer Programming}, 72\penalty0 (1--2):\penalty0
  3--21, 2008.

\bibitem[Balakrishnan et~al.(2009)Balakrishnan, Sankaranarayanan, Ivancic, and
  Gupta]{DBLP:conf/emsoft/BalakrishnanSIG09}
Gogul Balakrishnan, Sriram Sankaranarayanan, Franjo Ivancic, and Aarti Gupta.
\newblock Refining the control structure of loops using static analysis.
\newblock In \emph{EMSOFT}, pages 49--58. ACM, 2009.
\newblock \isbn{978-1-60558-627-4}.
\newblock \doi{10.1145/1629335.1629343}.

\bibitem[Biere et~al.(2009)Biere, Heule, van Maaren, and Walsh]{Handbook_SAT}
Armin Biere, Marijn Heule, Hans van Maaren, and Toby Walsh, editors.
\newblock \emph{Handbook of satisfiability}, volume 185 of \emph{Frontiers in
  Artificial Intelligence and Applications}.
\newblock IOS Press, Amsterdam, 2009.
\newblock \isbn{978-1-58603-929-5}.

\bibitem[Blanchet et~al.(2003)Blanchet, Cousot, Cousot, Feret, Mauborgne,
  Min\'e, Monniaux, and Rival]{ASTREE_PLDI03}
Bruno Blanchet, Patrick Cousot, Radhia Cousot, J\'er\^ome Feret, Laurent
  Mauborgne, Antoine Min\'e, David Monniaux, and Xavier Rival.
\newblock A static analyzer for large safety-critical software.
\newblock In \emph{Programming Language Design and Implementation (PLDI)},
  pages 196--207. ACM, 2003.
\newblock \isbn{1-58113-662-5}.
\newblock \doi{10.1145/781131.781153}.

\bibitem[Cousot and Cousot(1992)]{CousotCousot_JLC92}
Patrick Cousot and Radhia Cousot.
\newblock Abstract interpretation frameworks.
\newblock \emph{J. of Logic and Computation}, pages 511--547, August 1992.
\newblock \issn{0955-792X}.
\newblock \doi{10.1093/logcom/2.4.511}.

\bibitem[Cousot and Halbwachs(1978)]{CousotHalbwachs78}
Patrick Cousot and Nicolas Halbwachs.
\newblock Automatic discovery of linear restraints among variables of a
  program.
\newblock In \emph{Principles of Programming Languages (POPL)}, pages 84--96.
  ACM, 1978.
\newblock \doi{10.1145/512760.512770}.

\bibitem[Cousot et~al.(2005)Cousot, Cousot, Feret, Mauborgne, Min\'e, Monniaux,
  and Rival]{ASTREE_ESOP05}
Patrick Cousot, Radhia Cousot, J\'er\^ome Feret, Laurent Mauborgne, Antoine
  Min\'e, David Monniaux, and Xavier Rival.
\newblock The {ASTR\'EE} analyzer.
\newblock In \emph{Programming Languages and Systems (ESOP)}, number 3444 in
  Lecture Notes in Computer Science, pages 21--30. Springer Verlag, 2005.
\newblock \isbn{3-540-25435-8}.
\newblock \doi{10.1007/b107380}.

\bibitem[de~Moura and Bj{\o}rner(2008)]{DBLP:conf/tacas/MouraB08}
Leonardo~Mendon\c{c}a de~Moura and Nikolaj Bj{\o}rner.
\newblock {Z3}: An efficient {SMT} solver.
\newblock In \emph{TACAS}, volume 4963 of \emph{Lecture Notes in Computer
  Science}, pages 337--340. Springer, 2008.
\newblock \isbn{978-3-540-78799-0}.

\bibitem[Dutertre and de~Moura(2006)]{DBLP:conf/cav/DutertreM06}
Bruno Dutertre and Leonardo~Mendon\c{c}a de~Moura.
\newblock A fast linear-arithmetic solver for {DPLL(T)}.
\newblock In \emph{CAV}, volume 4144 of \emph{Lecture Notes in Computer
  Science}, pages 81--94. Springer, 2006.
\newblock \isbn{3-540-37406-X}.

\bibitem[Gawlitza and Monniaux(2011)]{Gawlitza_Monniaux_ESOP11}
Thomas Gawlitza and David Monniaux.
\newblock Improving strategies via {SMT} solving.
\newblock In \emph{ESOP}, number 6602 in Lecture Notes in Computer Science,
  pages 236--255. Springer Verlag, 2011.
\newblock \isbn{978-3-642-19717-8}.
\newblock \doi{10.1007/978-3-642-19718-5_13}.

\bibitem[Gonnord and Halbwachs(2006)]{DBLP:conf/sas/GonnordH06}
Laure Gonnord and Nicolas Halbwachs.
\newblock Combining widening and acceleration in linear relation analysis.
\newblock In \emph{Static analysis (SAS)}, volume 4134 of \emph{Lecture Notes
  in Computer Science}, pages 144--160. Springer Verlag, 2006.
\newblock \isbn{3-540-37756-5}.
\newblock \doi{10.1007/11823230_10}.

\bibitem[Gopan and Reps(2007)]{DBLP:conf/sas/GopanR07}
Denis Gopan and Thomas~W. Reps.
\newblock Guided static analysis.
\newblock In \emph{SAS}, volume 4634 of \emph{Lecture Notes in Computer
  Science}, pages 349--365. Springer, 2007.
\newblock \isbn{978-3-540-74060-5}.

\bibitem[Gulwani and Zuleger(2010)]{DBLP:conf/pldi/GulwaniZ10}
Sumit Gulwani and Florian Zuleger.
\newblock The reachability-bound problem.
\newblock In \emph{PLDI}, pages 292--304. ACM, 2010.
\newblock \isbn{978-1-4503-0019-3}.
\newblock \doi{10.1145/1806596.1806630}.

\bibitem[Halbwachs(1979)]{Halbwachs_PhD}
Nicolas Halbwachs.
\newblock \emph{D\'etermination automatique de relations lin\'eaires
  v\'erifi\'ees par les variables d'un programme}.
\newblock PhD thesis, Grenoble University, 1979.

\bibitem[Halbwachs et~al.(1997)Halbwachs, Proy, and Roumanoff]{Polka:FMSD:97}
Nicolas Halbwachs, Yann-Erick Proy, and Patrick Roumanoff.
\newblock Verification of real-time systems using linear relation analysis.
\newblock \emph{Formal Methods in System Design}, 11\penalty0 (2):\penalty0
  157--185, August 1997.

\bibitem[Harris et~al.(2010)Harris, Sankaranarayanan, Ivancic, and
  Gupta]{DBLP:conf/popl/HarrisSIG10}
William~R. Harris, Sriram Sankaranarayanan, Franjo Ivancic, and Aarti Gupta.
\newblock Program analysis via satisfiability modulo path programs.
\newblock In \emph{POPL}, pages 71--82. ACM, 2010.
\newblock \isbn{978-1-60558-479-9}.
\newblock \doi{10.1145/1706299.1706309}.

\bibitem[Henry(2011)]{julien-henry-m2r}
Julien Henry.
\newblock Static analysis by path focusing.
\newblock Master's thesis, Grenoble INP, 2011.
\newblock URL \url{http://www-verimag.imag.fr/~jhenry/pdf/M2R_report.pdf}.

\bibitem[Jeannet(2003)]{DBLP:journals/fmsd/Jeannet03}
Bertrand Jeannet.
\newblock Dynamic partitioning in linear relation analysis: Application to the
  verification of reactive systems.
\newblock \emph{Formal Methods in System Design}, 23\penalty0 (1):\penalty0
  5--37, 2003.

\bibitem[Jeannet and Min{\'e}(2009)]{DBLP:conf/cav/JeannetM09}
Bertrand Jeannet and Antoine Min{\'e}.
\newblock Apron: A library of numerical abstract domains for static analysis.
\newblock In \emph{CAV}, volume 5643 of \emph{Lecture Notes in Computer
  Science}, pages 661--667. Springer Verlag, 2009.
\newblock \isbn{978-3-642-02657-7}.
\newblock \doi{10.1007/978-3-642-02658-4_52}.

\bibitem[Kroening and Strichman(2008)]{Kroening_Strichman_08}
Daniel Kroening and Ofer Strichman.
\newblock \emph{Decision procedures}.
\newblock Springer Verlag, 2008.
\newblock \isbn{978-3-540-74104-6}.

\bibitem[Lattner and Adve(2004)]{Lattner:2004:LCF:977395.977673}
Chris Lattner and Vikram Adve.
\newblock {LLVM}: A compilation framework for lifelong program analysis \&
  transformation.
\newblock In \emph{CGO}, pages 75--86, Washington, DC, USA, August 2004. IEEE
  Computer Society.
\newblock \isbn{0-7695-2102-9}.
\newblock \doi{10.1109/CGO.2004.1281665}.

\bibitem[LLV(2011)]{LLVM_langref}
\emph{{LLVM} Language Reference Manual}.
\newblock LLVM team, 2011.
\newblock \url{http://llvm.org/docs/LangRef.html}.

\bibitem[Min{\'e}(2006)]{DBLP:journals/lisp/Mine06}
Antoine Min{\'e}.
\newblock The octagon abstract domain.
\newblock \emph{Higher-Order and Symbolic Computation}, 19\penalty0
  (1):\penalty0 31--100, 2006.
\newblock \doi{10.1007/s10990-006-8609-1}.

\bibitem[Monniaux and Bodin(2011)]{Monniaux_Bodin_APLAS11}
David Monniaux and Martin Bodin.
\newblock Modular abstractions of reactive nodes using disjunctive invariants.
\newblock In \emph{Programming Languages and Systems (APLAS)}, pages 19--33,
  2011.
\newblock \isbn{978-3-642-25317-1}.
\newblock \doi{10.1007/978-3-642-25318-8_5}.

\bibitem[Monniaux and Gonnord(2011)]{Monniaux_Gonnord_SAS11}
David Monniaux and Laure Gonnord.
\newblock Using bounded model checking to focus fixpoint iterations.
\newblock In \emph{Static analysis (SAS)}, volume 6887 of \emph{Lecture Notes
  in Computer Science}, pages 369--385. Springer Verlag, 2011.

\bibitem[Rival and Mauborgne(2007)]{Rival_Mauborgne_TOPLAS07}
Xavier Rival and Laurent Mauborgne.
\newblock The trace partitioning abstract domain.
\newblock \emph{Transactions on Programming Languages and Systems (TOPLAS)},
  29\penalty0 (5):\penalty0 26, 2007.
\newblock \issn{0164-0925}.
\newblock \doi{10.1145/1275497.1275501}.

\bibitem[Sharma et~al.(2011)Sharma, Dillig, Dillig, and
  Aiken]{DBLP:conf/cav/SharmaDDA11}
Rahul Sharma, Isil Dillig, Thomas Dillig, and Alex Aiken.
\newblock Simplifying loop invariant generation using splitter predicates.
\newblock In \emph{CAV}, volume 6806 of \emph{Lecture Notes in Computer
  Science}, pages 703--719. Springer Verlag, 2011.
\newblock \isbn{978-3-642-22109-5}.
\newblock \doi{10.1007/978-3-642-22110-1_57}.

\end{thebibliography}
\end{document}